\documentclass[twocolumn,english,aps,prb,superscriptaddress]{revtex4}
\usepackage[T1]{fontenc}
\usepackage[latin9]{inputenc}
\usepackage[a4paper]{geometry}
\usepackage{amsmath}
\usepackage{amssymb}
\usepackage{graphicx}
\usepackage{esint}

\makeatletter
\@ifundefined{textcolor}{}
{%
 \definecolor{BLACK}{gray}{0}
 \definecolor{WHITE}{gray}{1}
 \definecolor{RED}{rgb}{1,0,0}
 \definecolor{GREEN}{rgb}{0,1,0}
 \definecolor{BLUE}{rgb}{0,0,1}
 \definecolor{CYAN}{cmyk}{1,0,0,0}
 \definecolor{MAGENTA}{cmyk}{0,1,0,0}
 \definecolor{YELLOW}{cmyk}{0,0,1,0}
}

\usepackage{amsbsy}\usepackage{mciteplus}\usepackage{babel}

\usepackage{babel}

\makeatother

\usepackage{babel}
\begin{document}

\author{K. Kolasi\'{n}ski}

\affiliation{AGH University of Science and Technology, Faculty of Physics and
Applied Computer Science,\\
 al. Mickiewicza 30, 30-059 Kraków, Poland}

\author{B.Szafran}

\affiliation{AGH University of Science and Technology, Faculty of Physics and
Applied Computer Science,\\
 al. Mickiewicza 30, 30-059 Kraków, Poland}

\author{B. Hackens}

\affiliation{Université Catholique de Louvain (UCL), IMCN/NAPS, 2 Chemin du Cyclotron,
1348 Louvain-la-Neuve, Belgium}

\title{Multitip scanning gate microscopy for ballistic transport studies
in systems with two-dimensional electron gas}
\begin{abstract}
We consider conductance mapping of systems based on the two-dimensional
electron gas with scanning gate microscopy using two and more tips
of the atomic force microscope. The paper contains results of numerical
simulations for a model tip potential with a proposal of a few procedures
for extraction and manipulation the ballistic transport properties.
In particular, we demonstrate that the multi-tip techniques can be
used for readout of the Fermi wavelength, detection of potential defects,
filtering specific transverse modes, tuning the system into resonant
conditions under which a stable map of a local density of states can
be extracted from conductance maps using a third tip.
\end{abstract}
\maketitle

\section{Introduction}

Conductance ($G$) of open systems with the two-dimensional electron
gas (2DEG) in semiconductor heterostructures at low temperatures is
determined by scattering of the Fermi level electrons. Using the scanning
gate microscopy (SGM) technique\cite{sgmr} one probes the properties
of the electron transport in mesoscopic devices by variation of the
potential landscape for the Fermi level electrons with the charged
tip of the atomic force microscope (AFM) moving above the surface
of the sample. The potential of the charge at the tip is screened
by the two-dimensional electron gas \cite{kolasinski2013} which produces
a short-range form of the effective tip potential for the Fermi level
electrons. The SGM technique was widely used in the studies of the
ballistic electron transport, in particular to visualize the electron
trajectories as deflected by external magnetic field \cite{Aidala2000iles},
electron branching of the Coulomb flow \cite{topinka1,topinka2} including
an evidence of the quantum Braess paradox\cite{Pala}, formation of
a quantum ring potential with a controlled number of modes in each
arm\cite{Kozikov2014}, investigation of electron backscattering with
quantum point contacts\cite{topinka1,topinka2}, formation of Coulomb
islands \cite{Hackens2010} and mapping the local density of states
(LDOS)\cite{Pala2008,Sellier2010}. Besides the studies of the ballistic
flow, the tip potential is also used for the studies of the charge
flow in conditions of the Coulomb blockade in quantum dots \cite{cbm}.
The role of the tip is then to tune the chemical potential of the
confined electron system into the transport window defined by the
Fermi energies of the source and drain electrodes \cite{cbm}.

In this paper we consider possible applications of the SGM technique
for the studies of the ballistic transport using two or more tips
instead of a single one. We show that the double tip system can be
used to: i) measuring the Fermi wavelength, ii) mapping the potential
defects in the channel iii) mode filtering, iv) detection of localized
resonances in an experimental implementation of a stabiliation method,
v) tuning the system to resonant conditions when LDOS can be read-out
with a third tip.

Usage of several probes was implemented a few years ago for the scanning
tunneling microscopy (STM) studies of the sample surface. A version
of the STM technique using several independent tips was used in particular
to perform the four point measurements \cite{4p} for determination
of the surface properties.

\section{Model}

We consider the ballistic transport at the Fermi level electrons in
systems based on a two dimensional electron gas (2DEG). We neglect
the electron-electron interactions and we use the effective mass Hamiltonian
of form
\begin{equation}
\left\{ -\frac{\hbar^{2}}{2m_{\mathrm{eff}}}\nabla^{2}+V_{ext}(x,y)\right\} \psi(x,y)=E_{\mathrm{F}}\psi(x,y),\label{eqschrodinger-1}
\end{equation}
where $m_{\mathrm{eff}}=0.067m_{\mathrm{0}}$ is the effective mass
of GaAs, $E_{\mathrm{F}}$ is the Fermi level energy, and $V_{\mathrm{ext}}$
contains all external sources of the electrostatic potential (eg.
potential of the tips $V_{\mathrm{tip}}$). We assume that the potential
of the tip is given by short-range Lorentzian potential of amplitude
$U_{\mathrm{tip}}$ and width $w_{\mathrm{tip}}$
\begin{equation}
V_{\mathrm{tip}}(x,y)=\frac{U_{\mathrm{tip}}}{1+\left[\left(x-x_{\mathrm{tip}}\right)^{2}+\left(y-y_{\mathrm{tip}}\right)^{2}\right]/w_{\mathrm{tip}}^{2}}.\label{eqlorentz-1}
\end{equation}
This form of the potential was evaluated in previous \cite{kolasinski2013},
self-consistent Schrödinger-Poisson calculation. In the following
we assume that all the tip potentials have the same width and amplitude
and that the distance between them can be changed. The shape of the
discussed devices - tailored from the sample containing a two-dimensional
electron gas is described by hard-wall boundary conditions.

In order to solve the scattering problem we use a finite difference
implementation of the Transparent Boundary Method (TBM) \cite{Lent90,Lent94,Kolasinski2014QHI}.
For the boundary conditions in the input lead we use the standard
approach with the wave function given by superposition of incoming
and outgoing (reflected) transverse modes
\begin{eqnarray}
\psi_{\mathrm{input}}(x,y) & = & \sum_{k=1}^{M_{\mathrm{input}}}\left\{ a_{k}e^{ikx}\chi_{k}^{\mathrm{input}}(y)\right.\label{eq:phi_input}\\
 & + & \left.r_{k}e^{-ikx}\chi_{-k}^{\mathrm{input}}(y)\right\} ,\nonumber
\end{eqnarray}
where $M_{\mathrm{input}}$ is the number of current propagating transverse
modes $\chi_{k}^{\mathrm{input}}$ in the input lead, $a_{k}$ and
$r_{k}$ are the incoming and reflection amplitudes. For the output
lead we assume that the wave function is given by formula
\begin{eqnarray}
\psi_{\mathrm{output}}(x,y) & = & \sum_{k=1}^{M_{\mathrm{output}}}t_{k}e^{ikx}\chi_{k}^{\mathrm{\mathrm{output}}}(y),\label{eq:phi_output}
\end{eqnarray}
where $M_{\mathrm{output}}$ is the number of transverse modes $\chi_{k}^{\mathrm{output}}$
in the output lead and $t_{k}$ is the outgoing amplitude. The transverse
modes were calculated with the method presented in Ref. \cite{Zwierzycki2008}.
Matching the boundary conditions \eqref{eq:phi_input} and \eqref{eq:phi_output}
with the wave function calculated inside the device one finds the
solution for the scattering problem. After solution of the Eq. \eqref{eqschrodinger-1}
(for details see \cite{Kolasinski2014QHI}) for each incoming mode
$i$ the conductance of the system is calculated from the transmission
probability $T_{i}$ using the zero-temperature Landauer formalism
\[
G=G_{0}\sum_{i=1}^{M_{\mathrm{input}}}T_{i},
\]
with $G_{0}=\frac{2e^{2}}{h}.$

In all the cases considered below we assume that both leads have the
same width, thus $M\equiv M_{\mathrm{input}}=M_{\mathrm{output}}$.
We choose the discretization grid $\Delta x=\Delta y=4$nm. In further
discussion we will refer to the local density of states which is defined
as the sum of electron densities incoming from the left and right
lead \cite{kolasinski2013}.

\section{Impurities mapping}

\begin{figure}[h]
\begin{centering}
\includegraphics[width=0.4\paperwidth]{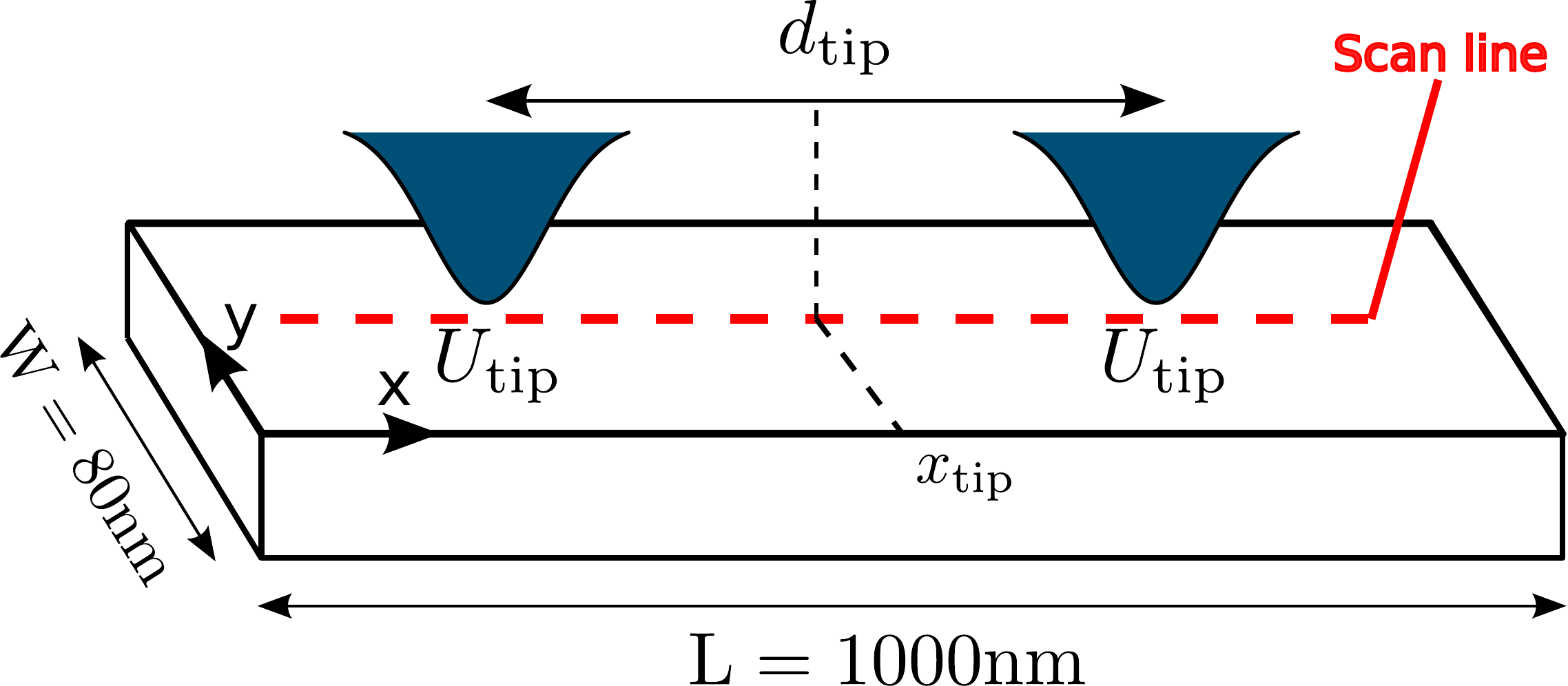}
\par\end{centering}

\caption{\label{fig:kanalSchemat}Sketch of the system considered in Sections
\ref{s1} and \ref{s2}. We consider a channel of width $W=80$ nm
and the system of two tips with the same amplitude $U_{\mathrm{tip}}=5$meV
and width $w_{\mathrm{tip}}=15$nm The length of the computational
channel is $L=1000$nm. The tips are above the axis of the channel
(red dashed line) and are separated by a distance $d_{\mathrm{tip}}$.
The center of the system of two tips is located at position $x_{\mathrm{tip}}$. }
\end{figure}

\begin{figure}[h]
\begin{centering}
\includegraphics[width=0.4\paperwidth]{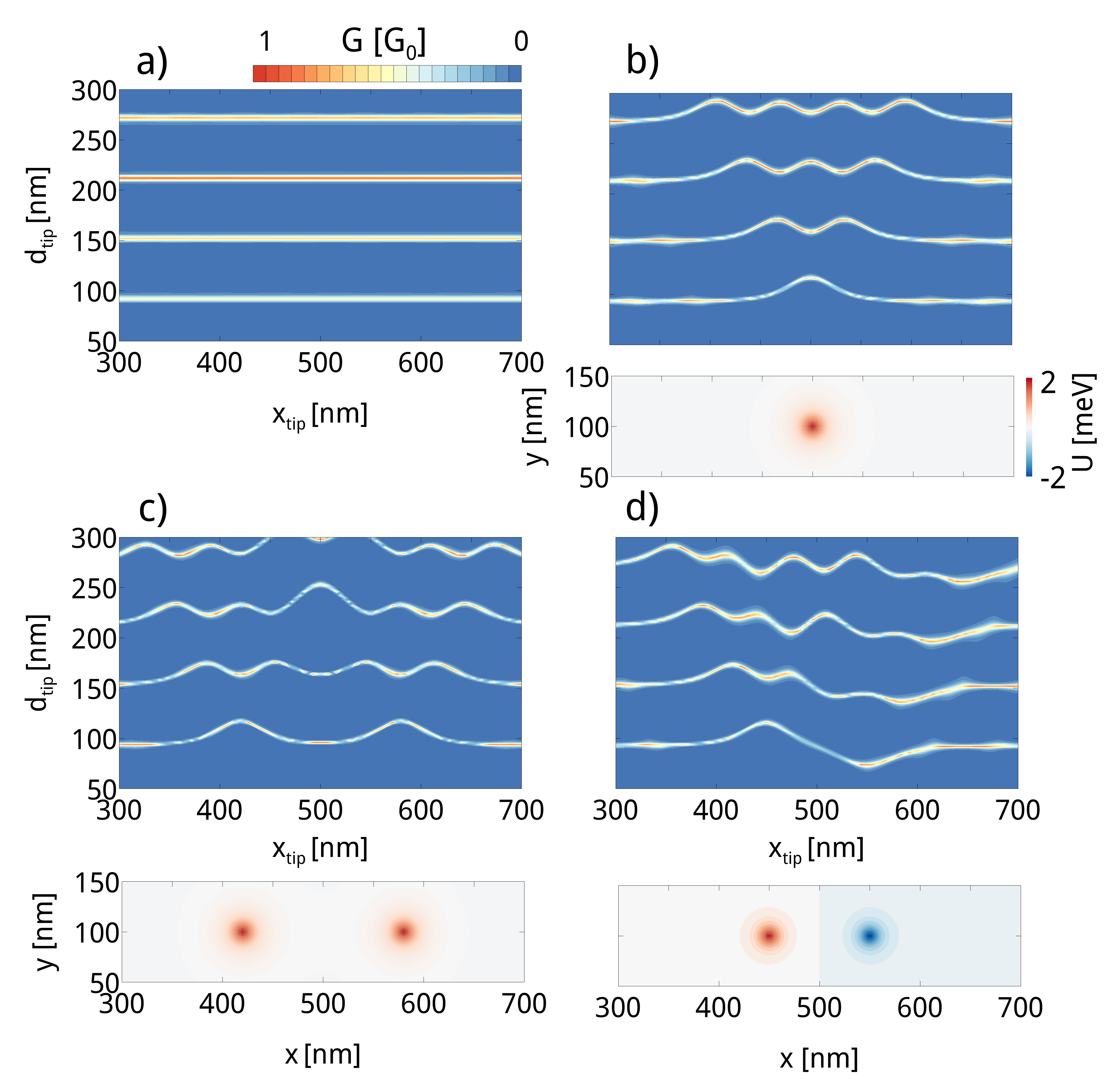}
\par\end{centering}

\caption{\label{fig:skanyKanal}Results for $E_{\mathrm{F}}=2.5$meV (one band
transport) a) The DT scan obtained for a clean channel. b) Added impurity
in the middle of channel with amplitude $U_{\mathrm{imp}}=2$meV and
of width $w_{\mathrm{imp}}=10$nm modeled with Eq. \eqref{eqlorentz-1}.
c) the same but for two impurities separated by distance 160nm and
d) obtained for two impurities of different sign $\pm2$meV separated
by distance 100nm. Plots below show the potential profile in the channel
for each case (b-d). }
\end{figure}

Let us start our discussion with the device presented in Fig. \ref{fig:kanalSchemat}.
We consider a long ($1000$nm) and narrow ($80$nm) channel. 
For the Fermi energy $E_{\mathrm{F}}=2.5$meV there is only one current-propagating
transverse mode in the channel ($M=1$). We consider a double tip
system assuming that tips move along the $x$ axis. The center of
the system of the two tips is denoted by $x_{\mathrm{tip}}$ (see
Fig. \ref{fig:kanalSchemat}). Both tips are separated by a distance
of $d_{\mathrm{tip}}$. We assume $U_{\mathrm{tip}}=5$meV and $w_{\mathrm{tip}}=15$nm,
For a \textit{single} tip with these parameters the conductance of
the system is reduced to $G=0.12G_{0}$. 
In Fig. \ref{fig:skanyKanal}(a) we show the result for the conductance
$G$ of the system as a function of $x_{tip}$ and the intertip distance
$d_{\mathrm{tip}}$ for a clean channel. The $d_{\mathrm{tip}}$ dependence
of conductance reveals a series of resonances separated by distance
$\lambda_{F}/2\approx60$nm, which are related to formation of standing
waves between the tips. Thus with the two tips one can probe the Fermi
wavelength.

Let us now consider the system with a single potential defect present
in the middle of channel. The defect is modeled with Eq. \eqref{eqlorentz-1}
with $U_{\mathrm{imp}}=2$meV and $w_{\mathrm{imp}}=10$nm (see the
potential profile below the Fig. \ref{fig:skanyKanal}(b)). Now the
SGM image reveals new features - the resonance lines bend and oscillate
in the vicinity of the impurity. Far from the impurity the resonance
lines return to the same position as in Fig. \ref{fig:skanyKanal}(a).
Note that the first resonance line (around the $d_{\mathrm{tip}}\approx90$nm)
resemble the potential profile inside the channel. In Fig. \ref{fig:skanyKanal}(c,d)
we show that the first resonance line follows the potential profile
of the defect. This effect can be explained in terms of the semi-classical
WKB approximation. For slowly varying potential $U(x)$ and $E_{\mathrm{F}}>U(x)$
the wave vector of propagating wave function $e^{\pm ik_{\mathrm{F}}x}$
is given by $k_{\mathrm{F}}=\sqrt{2m_{\mathrm{eff}}\left(E_{\mathrm{F}}-U(x)\right)}/\hbar$.
Now if $U(x)>0$ the phase of the wave propagating through the potential
hill will be delayed by some value $\Delta\psi$ in comparison to
the phase of the unperturbed system. Thus in order to restore the
resonance for standing wave one must increase the distance between
the tips, by the value which compensates for the delayed $\Delta\psi$
phase. For $U(x)<0$ the distance between the DT has to be decreased
(see Fig. \ref{fig:skanyKanal}(d)).

The system of a single tip and a repulsive defect is equivalent
to the double-tip system, hence the position of the defect can be resolved by a scan
of a single-tip by a shift of the energy lines. Nevertheless,
in order to extract the profile of the potential of a defect one needs two tips and not a single one.

\label{s1}

\begin{figure}[h]
\begin{centering}
\includegraphics[width=0.4\paperwidth]{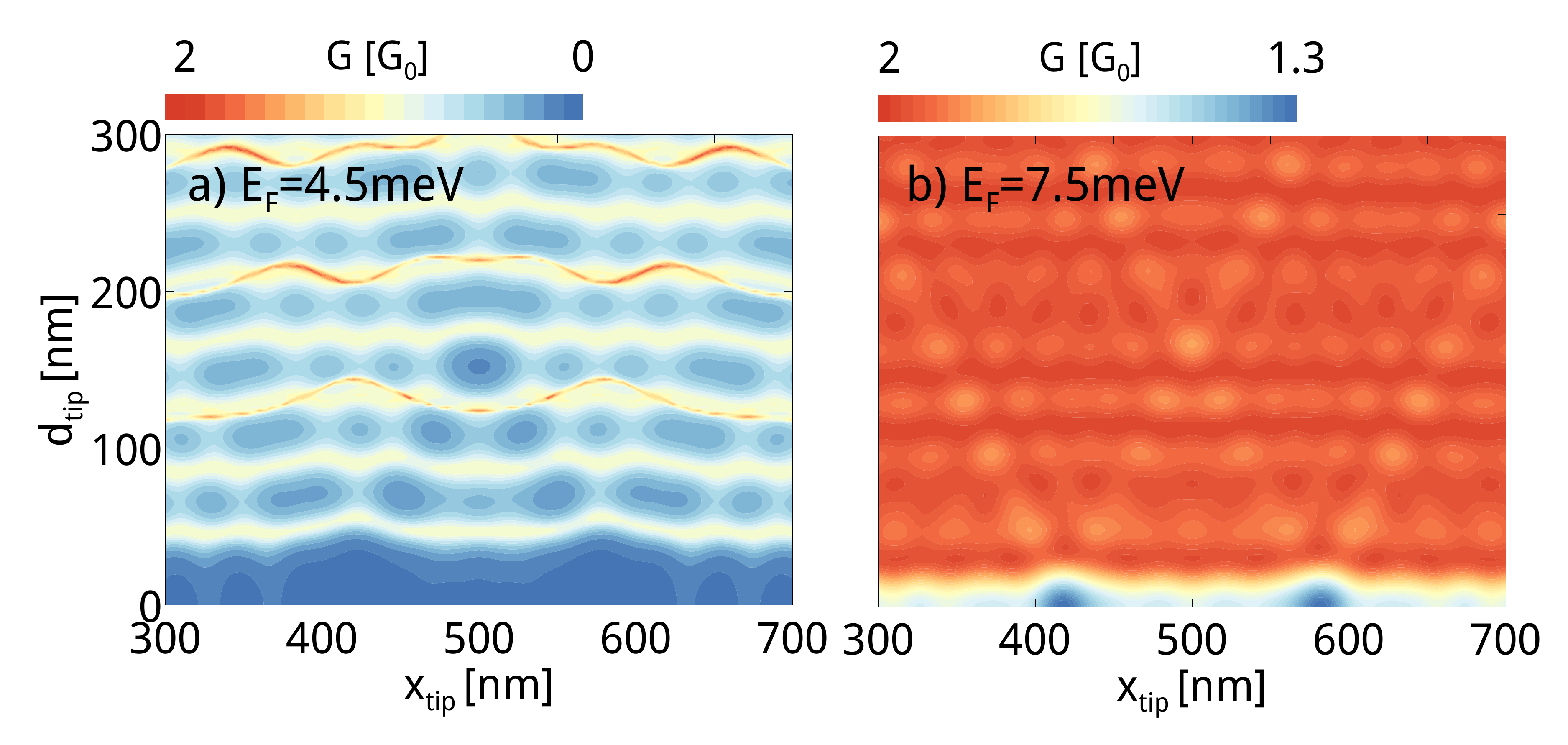}
\par\end{centering}

\caption{\label{fig:skanKanal2}The same as in Fig. \ref{fig:skanyKanal}(c)
but for higher Fermi energies. a) Results for $E_{\mathrm{F}}=4.5$meV
(two band transport) and b) $E_{\mathrm{F}}=7.5$meV.}
\end{figure}

In Fig. \ref{fig:skanKanal2} we show the results for the channel
with two impurities (same as in Fig. \ref{fig:skanyKanal}(c)) but
for higher energies $E_{\mathrm{F}}$. At higher energies {[}Fig.
\ref{fig:skanKanal2}(a){]} the resolution of the images is reduced
with the resonance lines getting closer and wider, but the image still
allows one to map the defect potential distribution along the channel.
Note that in Fig. \ref{fig:skanKanal2}(a) we notice the resonances
of Fig. \ref{fig:skanyKanal}(c) for both the lowest subband (resonance
sequence starting near $d_{tip}\simeq30$ nm) and the second transverse
mode (the sequence starts for $d_{tip}\simeq100$ nm). For $E_{\mathrm{F}}=7.5$meV
(see Fig. \ref{fig:skanKanal2}(b)) the image looses the features
visible in Figs. \ref{fig:skanyKanal}, but the $W$ shaped line clearly
indicates the position of the defects.

\begin{figure}[h]
\includegraphics[width=0.4\paperwidth]{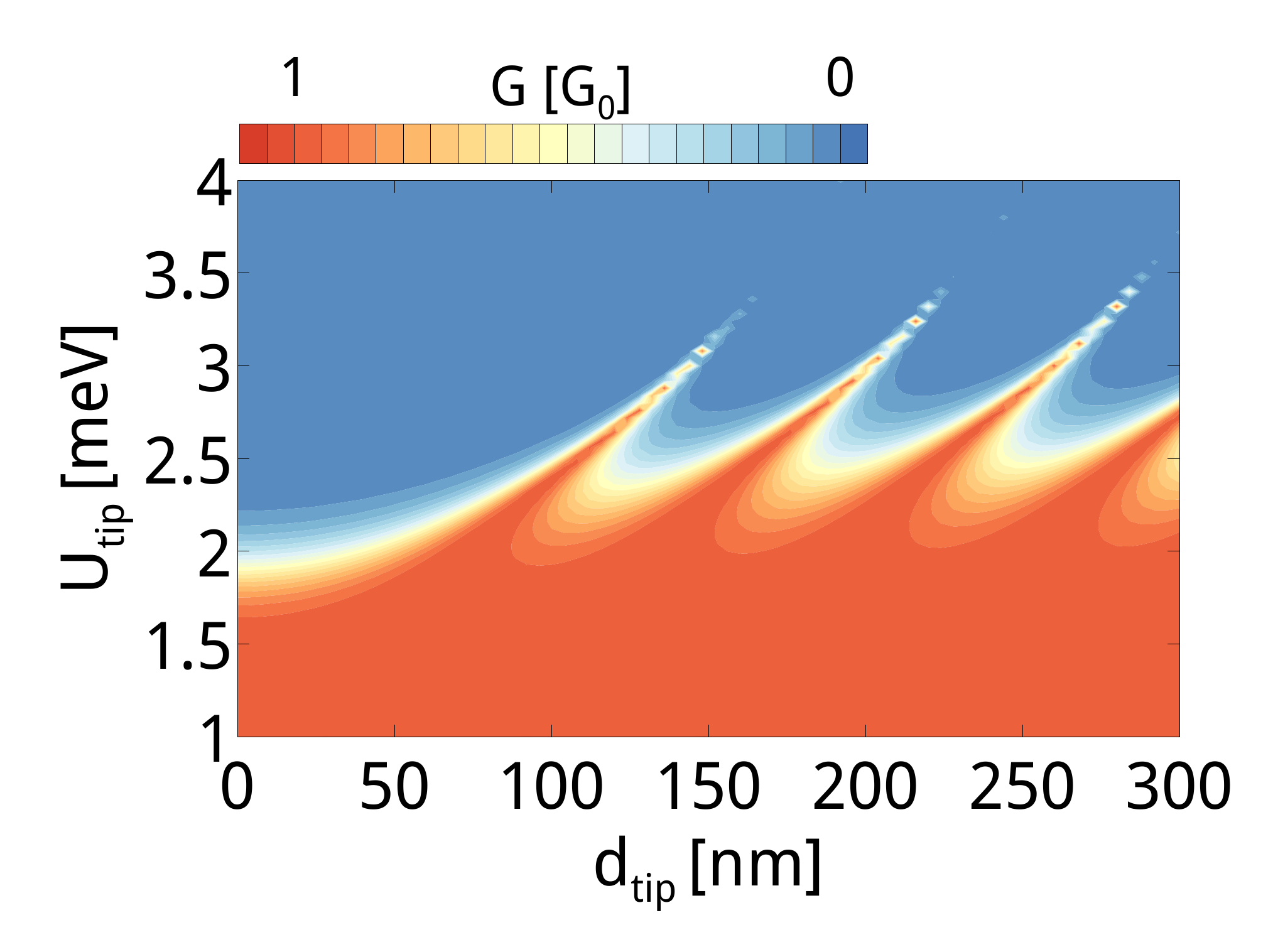} \caption{Conductance of the system as a function of the tip amplitude $U_{\mathrm{tip}}$
and distance $d_{\mathrm{tip}}$ between the tips obtained for $E_{\mathrm{F}}=2.5$meV
and width of the tip potential $w_{\mathrm{tip}}=50$nm.}

\label{fig:skanUtipDelta}
\end{figure}

For completeness we considered a clean channel with the width of the
tip potential increased to $w_{\mathrm{tip}}=50$nm. In Fig. \ref{fig:skanUtipDelta}
we plotted the conductance as a function of the intertip distance
$d_{tip}$ and the height of the tip potential. For $U_{tip}=2$ meV
the system is transparent for the electron flow, formation of the
resonances as discussed above for $w_{tip}=15$ nm appear higher in
the energy and for $U_{tip}>3.5$ meV the ballistic electron flow
is blocked. For larger values of $U_{tip}$ formation of a quantum
dot supported by the tips should be expected with the transport dominated
by the Coulomb blockade, which is however, outside the scope of the
present work. The best resolution of the scan as a function of $d_{tip}$
is obtained just below the cutoff of the ballistic transport.

To conclude this Section, we find that double tip system can be used
to read out the Fermi wavelength and the potential profile along the
channel.

\section{Transverse mode filter}

\label{s2} 
\begin{figure}[h]
\begin{centering}
\includegraphics[width=0.4\paperwidth]{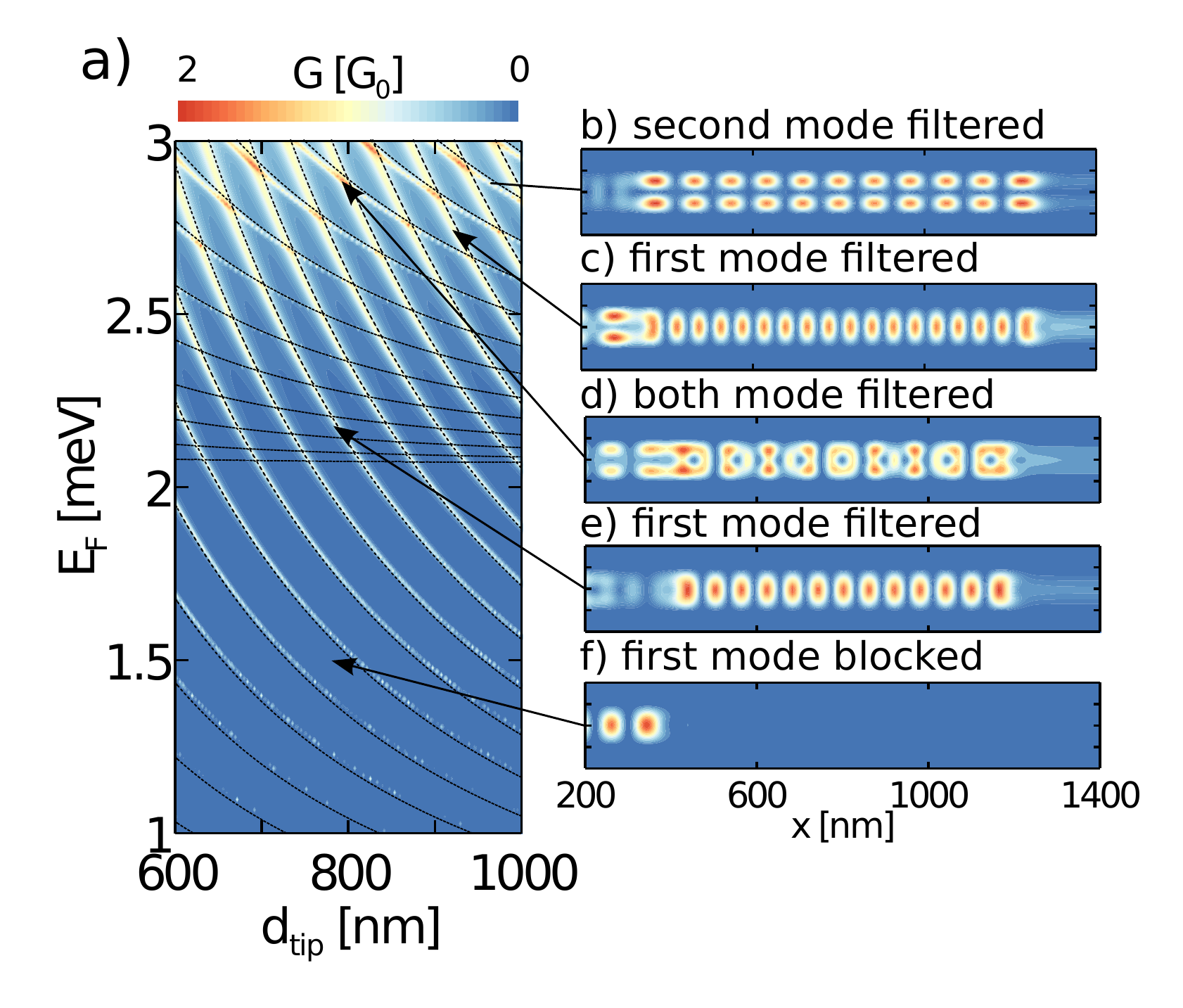}
\par\end{centering}

\caption{\label{fig:filter}a) The conductance $G$ of the long channel of
width $W=100$nm as a function of $d_{\mathrm{tip}}$ and $E_{\mathrm{F}}$.
Dashed lines deposited to the (a) shows the energy spectrum a two-dimensional
quantum well of length $d_{\mathrm{tip}}-2w_{\mathrm{tip}}$. The
shift by $2w_{\mathrm{tip}}$ accounts for the finite size of the
tip potential. (b-f) Electron density $|\psi|^{2}$ obtained for values
of ($d_{\mathrm{tip}}$,$E_{\mathrm{F}}$) pointed by arrows.}
\end{figure}

In Fig. \ref{fig:skanKanal2}(a) we observed the resonances
of the first and second subband (the red lines near $d_{tip}=100$,
200 and 300 nm ). These resonances occur separately or in overlap
with the lines of the lowest resonance depending on the values of
$d_{\mathrm{tip}}$ and $x_{\mathrm{tip}}$. Thus choosing a proper
value of $d_{\mathrm{tip}}$ and $x_{\mathrm{tip}}$ one can make
a device which will filter out specific modes. For the proof of principle,
we will restrict our consideration to the first and second mode. The
resonance lines are plotted in Fig. \ref{fig:filter}(a) as functions
of the distance between the tips and the Fermi energy for the clean
channel (as in Fig. \ref{fig:kanalSchemat}) but with the width increased
to $W=100$nm. At low Fermi energies only the first transverse mode
is present in the conductance, and the double tip stops the transport
unless the Fermi energy coincides with the resonances localized between
the tips. For higher values of $E_{\mathrm{F}}$ the resonance lines
of the second mode appear and at lines in the ($d_{\mathrm{tip}}$,$E_{\mathrm{F}}$)
plane which intersects with the first mode lines. The images (b-f)
in Fig. \ref{fig:filter} show the electron density for work points
marked by arrows in Fig. \ref{fig:filter}(a). We can see that for
a given energy $E_{\mathrm{F}}$ one may filter out a specific transverse
mode, by changing the $d_{\mathrm{tip}}$ distance.

In Fig. \ref{fig:filter}(a) we plotted the energy spectrum of an infinite quantum
well of width 100 nm and length $d_{tip}-2w_{tip}$. The conductance resonances follow exactly
the energy spectrum for a infinite quantum well.

\section{Detection of localized resonances}

\begin{figure}[h]
\begin{centering}
\includegraphics[width=0.4\paperwidth]{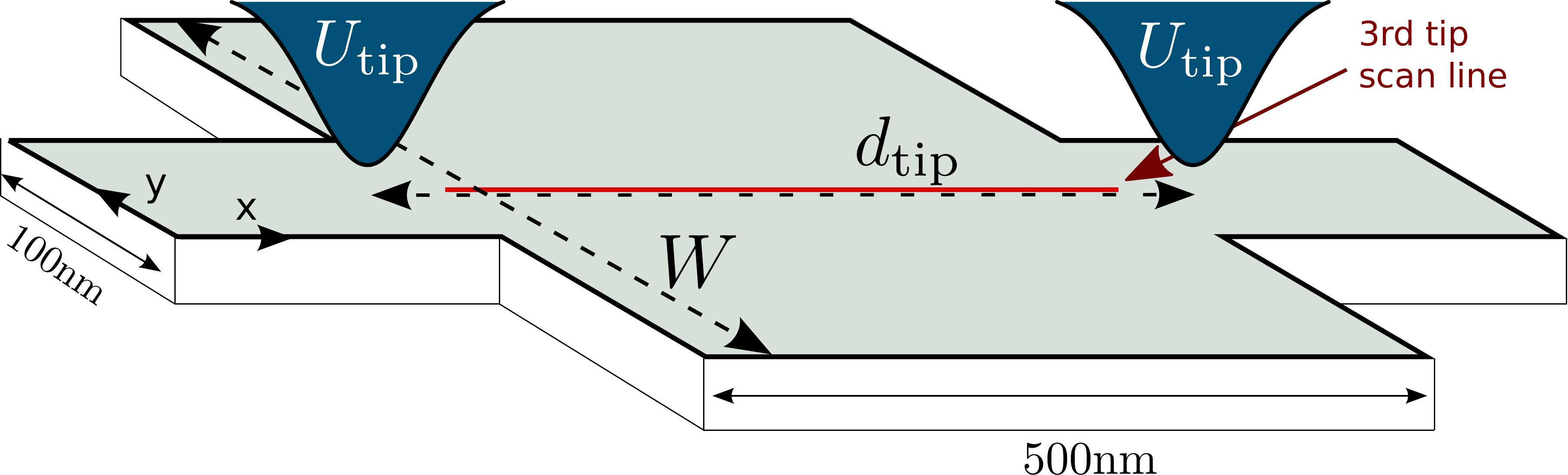}
\par\end{centering}

\caption{\label{fig:wnekaSchemat}Sketch of the second system considered in
this paper. Long channel of width 100nm is coupled to the resonant
cavity of variable width $W$ and length 500nm. The same as before
we have two tips in the system one in the left lead and second in
the right lead. Both tips have the same parameters of $U_{\mathrm{tip}}=5$meV
and $w_{\mathrm{tip}}=15$nm and are separated by distance $d_{\mathrm{tip}}$.
Red line corresponds to the third tip scan discussed in the last paragraph.}
\end{figure}

\begin{figure}[h]
\begin{centering}
\includegraphics[width=0.3\paperwidth]{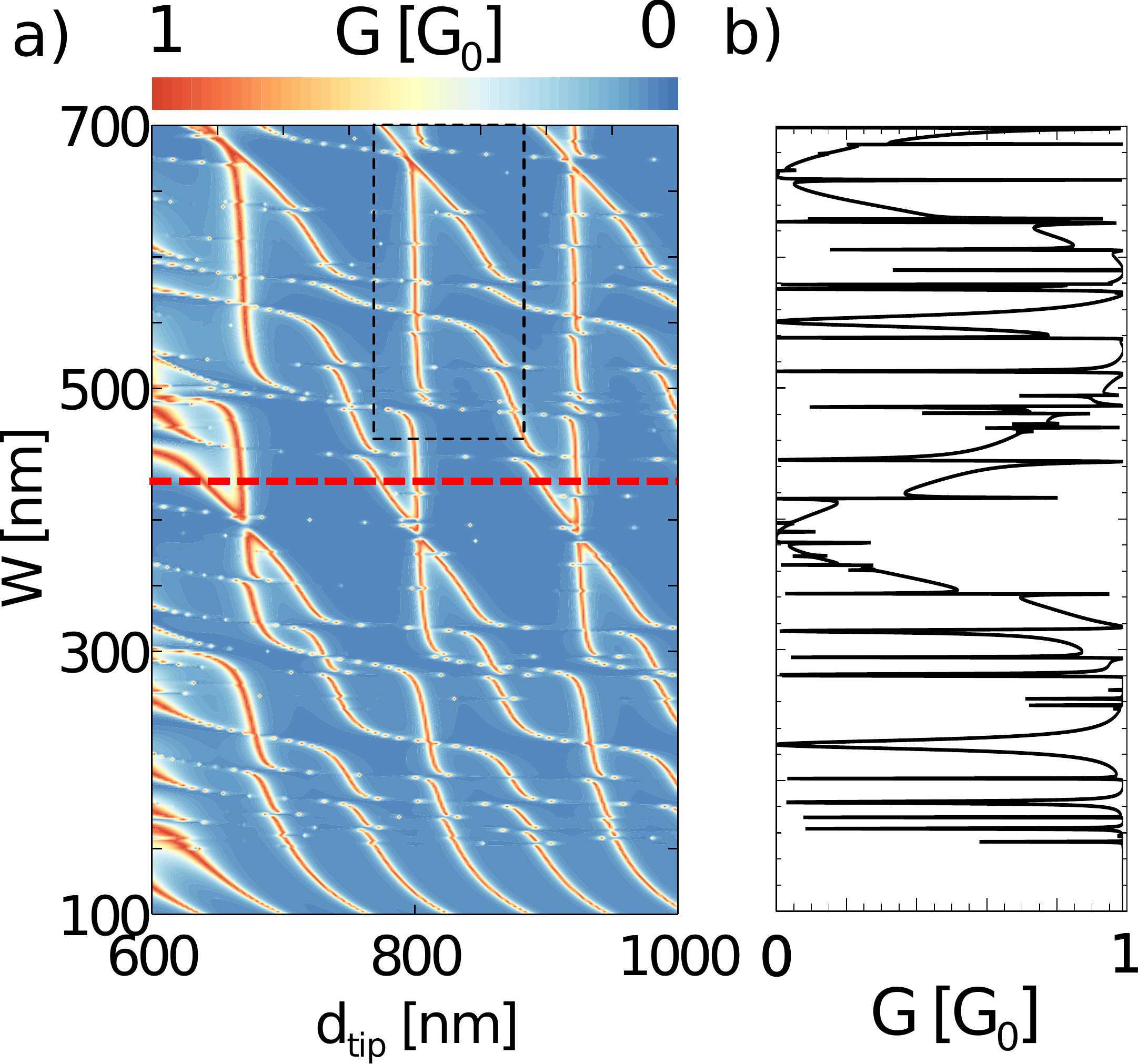}
\par\end{centering}

\caption{\label{fig:rezonanseWtip}a) The conductance of the system depicted
in Fig. \ref{fig:wnekaSchemat} as a function of $d_{\mathrm{tip}}$
and cavity width $W$. Dashed rectangle corresponds to the zoomed
area shown in Fig. \ref{fig:rezonanseZumy}(a). Red line corresponds
to scan with third tip discussed in the next paragraph. (b) Conductance
of the system presented in Fig. \ref{fig:wnekaSchemat} as a function
of $W$ but without DT. }
\end{figure}

Let us now consider the system depicted in Fig. \ref{fig:wnekaSchemat}
with a channel of width 100nm connected to the quantum cavity of length
500nm and width $W$. In order to get the best resolved images we
set the $E_{\mathrm{F}}=2$meV (single subband transport within the
channel). In Fig. \ref{fig:rezonanseWtip}(a) we show the conductance
of the system as a function of the $d_{\mathrm{tip}}$ and width $W$
of the cavity. We kept the center of the double tip fixed in the middle
of the resonant cavity. For small values of $W$ around $100$nm the
resonance lines are very similar to those of Fig. \ref{fig:filter}(a).
For a changed width of the cavity the cavity-localized energy levels
vary as $\propto W^{-2}$. Thus changing the width $W$ we should
expect behavior of the resonance lines similar to the ones of Fig.
\ref{fig:filter}(a). Large values of $W$ lead to a complex behavior
of the resonance lines but with well distinguishable patterns of X-shaped
lines which appear with a period of $\lambda_{F}$ along $d_{\mathrm{tip}}$
axis.

In Fig. \ref{fig:rezonanseWtip}(b) we show the conductance of the
system without the tips. The conductance contains a series of sharp
resonances of a Fano type corresponding to quasi-bound states localized
within the cavity. Most of the resonances visible in Fig. \ref{fig:rezonanseWtip}(b)
are also present in Fig. \ref{fig:rezonanseWtip}(a) as nearly horizontal
lines -- independent of the distance between the tips -- which suggests
that they correspond to the quasi-localized stated of the cavity.
The nearly horizontal lines are very thin indicating a long lifetime
of the resonances. The lines with a steeper dependence on $d_{tip}$
in Fig. \ref{fig:rezonanseWtip}(a) correspond to resonances supported
by the tips. The lines are wide - indicating a stronger coupling to
the channel. A study probing the conductance as a function of the
distance between the tips would be an experimental implementation
of the stabilization methods \cite{stb} for detection of localized
resonances.

\begin{figure}[h]
\begin{centering}
\includegraphics[width=0.4\paperwidth]{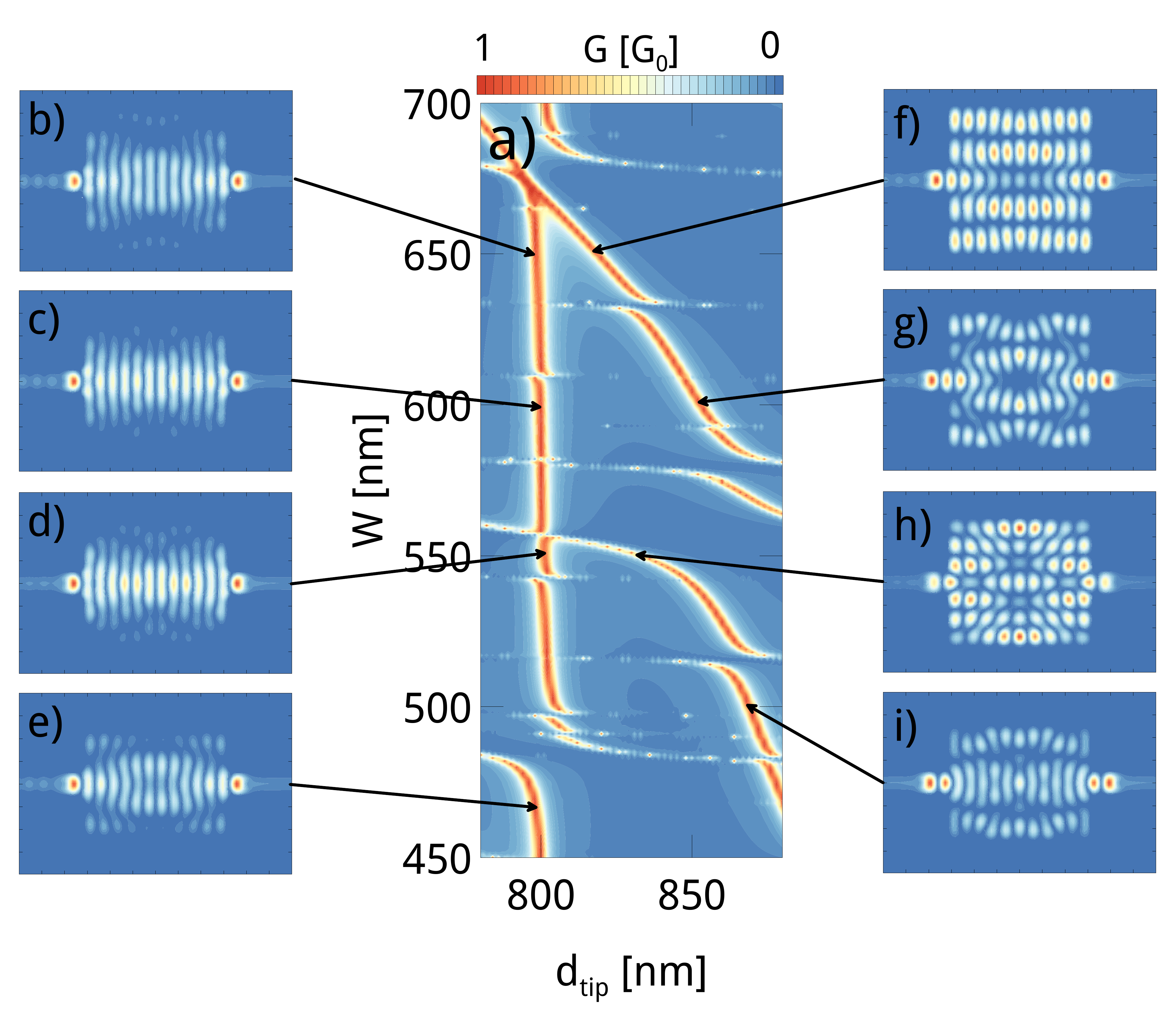}
\par\end{centering}

\caption{\label{fig:rezonanseZumy}(a) Zoom of figure \ref{fig:rezonanseWtip}(a)
marked by the dashed rectangle. (b-i) Probability density for the
the electron incoming from the left for $d_{\mathrm{tip}}$ and $W$
pointed by the arrows. (b-e) Antisymmetric resonances (f-i) symmetric
resonances.}
\end{figure}

The zoom of region of the Fig. \ref{fig:rezonanseWtip} marked by
the black rectangle is depicted in Fig. \ref{fig:rezonanseZumy}(a).
In this picture we can distinguish three types of resonance lines:
a) vertical -- almost independent of the width $W$ of the cavity,
b) resonances which vary with $W$ and c) well visible horizontal
resonances -- independent of the distance between DT which correspond
to the quasi localized state in the cavity. We found in general that
the difference between the two first types of resonances come from
the symmetry of the resonant scattering densities. The scattering
densities for lines of type a) and b) are symmetric with respect to
the center of the cavity in the $x$ direction. The resonances that
are independent of $W$ have a nodal surface at the symmetry axis,
while the other have a maximum on the symmetry line. The former are
strongly localized in the center of cavity (see Figs. \ref{fig:rezonanseZumy}(b-e))
and the latter form resonances which are delocalized over the entire
cavity (see Figs. \ref{fig:rezonanseZumy}(f-g,i)). The delocalized
resonances react stronger to the value of $W$ in Fig. \ref{fig:rezonanseWtip}(a).
Fig. \ref{fig:rezonanseZumy}(h) shows a resonance of a third type
- of the energy that is weakly dependent of $d_{tip}$ -- which corresponds
to the resonances supported by the cavity itself.

\begin{figure}[h]
\begin{centering}
\includegraphics[width=0.3\paperwidth]{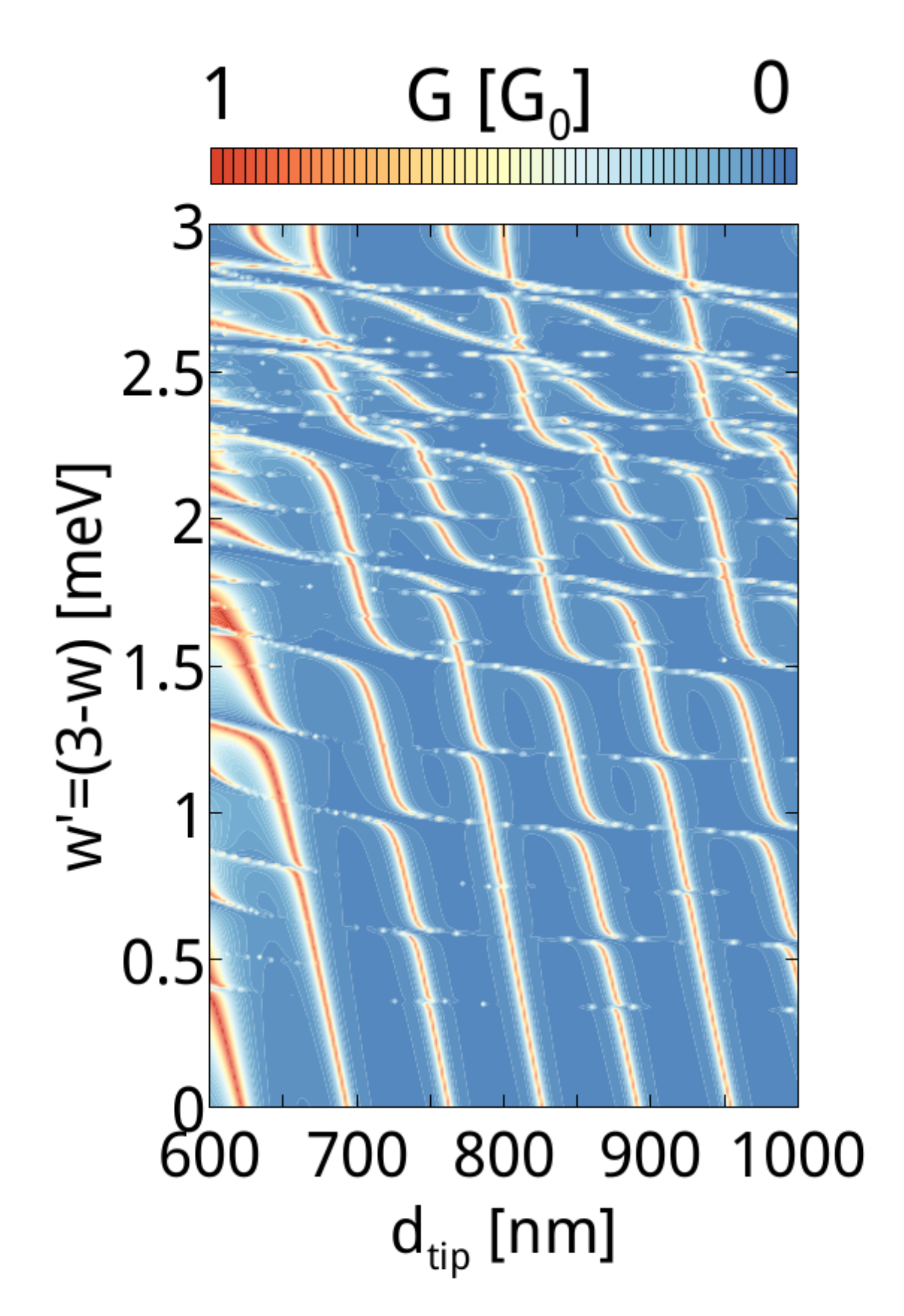}
\par\end{centering}

\caption{\label{fig:parabola}The conductance of the system depicted in Fig.
\ref{fig:wnekaSchemat} as a function of $d_{\mathrm{tip}}$ and angular
frequency $\omega$ of the oscillator. The parametrization $\omega'=3\mathrm{meV-}\omega$
is applied in order to keep the same interpretation of $y$ axis (width
of the cavity) as in Fig. \ref{fig:rezonanseWtip}(a).}
\end{figure}

In experiments variation of the geometry of the cavity can be accomplished
by tuning an electrostatic confinement in the $y$ direction due to
external gating. Since the electrostatic confinement potential is
usually parabolic at its origin we replaced the hard wall confinement
by
\begin{equation}
U_{\mathrm{parabolic}}=\frac{1}{2}\omega'^{2}y^{2}=\frac{1}{2}(3\mathrm{meV}-\omega)^{2}y^{2}.\label{eq:parabola}
\end{equation}
With this potential the effective width of the cavity is controlled
by the harmonic oscillator energy. The parametrization was chosen
in order to get better comparison with Fig. \ref{fig:rezonanseWtip}.
The value $\omega=3\mathrm{meV}\Rightarrow\omega'=0$meV corresponds
to a small value of $W$ in Fig. \ref{fig:wnekaSchemat} and channel-like
behavior of conductance. Value $\omega=0$ meV$\Rightarrow\omega'=3$meV
corresponds to the large values of $W$. Comparing the images Fig.
\ref{fig:rezonanseWtip} and Fig. \ref{fig:parabola} we can see that
both share the same features: a) X-shaped resonance lines are present,
b) some resonances depend weakly on $W$ as a function of the cavity
width (antisymmetric ones) and some vary with $d_{tip}$ (symmetric
states), c) horizontal resonance lines are also present. Note that
in this case we do not find the vertical resonances in the conductance
image. This is because the variation of $\omega$ changes the potential
profile in whole cavity and not only the width of the cavity (as it
was in the previous example).

Concluding this Section, we found that using double tip system with
the cavity of variable width one may tune the system to the specific
resonance i) symmetric or antisymmetric resonances supported between
the tips ii) or resonances quasi-localized within the cavity (horizontal
lines in Figs. \ref{fig:rezonanseWtip} and \ref{fig:parabola}).

\section{Mapping the local density at Fermi level}

Let us now discuss the possibility of mapping the LDOS inside the
device presented in Fig. \ref{fig:wnekaSchemat}. In the two-dimensional
systems the conductance maps can be correlated to the local density
of states for weak tip perturbation when the electron scattering wave
functions incoming from the left and right leads are the same up to
a constant phase. The latter case can be supported \cite{KolasinskiNJP}
by the perturbation theory \cite{Jalabert2010,Gorini2013} for which
the first-order correction to the conductance is simply proportional
to the LDOS. The scattering densities for the electron incoming from
the left lead plotted in Fig. \ref{fig:rezonanseZumy}(b-i) are highly
symmetric since they correspond to the resonances induced by the double
tip system in a symmetric cavity. Thus for inverted current direction
the scattering density inside the cavity stays the same and this should
lead good correlation between LDOS and corresponding $G$ map. In
general we find \cite{kolasinski2013,KolasinskiNJP} that the correlation
LDOS-G is well resolved near the Fano resonances.

In the following calculation we assume that the distance between the
two tips placed within the channel is set to a resonance, and the
third tip scans the surface of the cavity for detection of the resonant
LDOS. The assumed third tip parameters are $U_{\mathrm{tip}}=1$meV
and $w_{\mathrm{tip}}=10$nm. We need a pronounced variation of $G$
as a function of the tip position and the adopted height of the tip
potential is quite large as compared to the Fermi energy kept at 2
meV.

\begin{figure*}
\begin{centering}
\includegraphics[width=0.7\paperwidth]{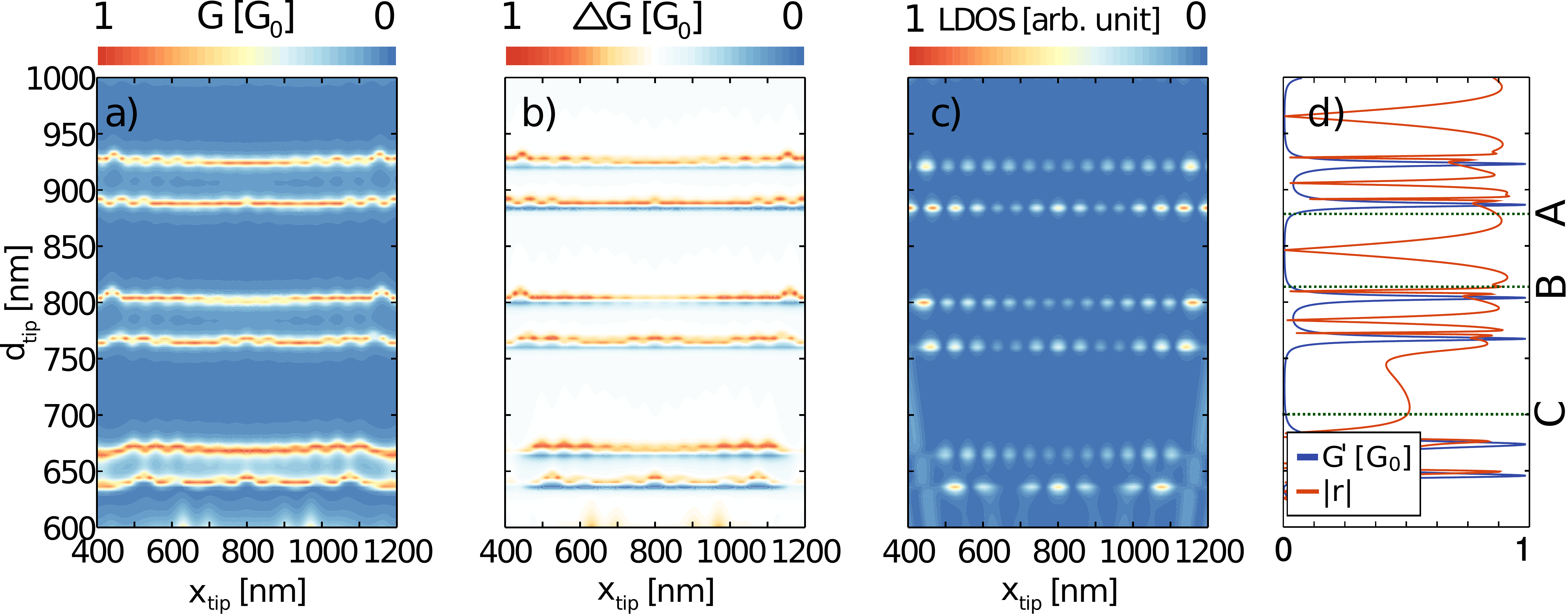}
\par\end{centering}

\caption{\label{fig:korelacje}a) Conductance $G$ of the system as a function
of third tip position $x_{\mathrm{tip}}$(moved along the red line
in Fig. \ref{fig:wnekaSchemat}) and distance between the two tips
(red line in Fig. \ref{fig:rezonanseWtip}(a)) b) The same but $\Delta G=G-G'$
is plotted, where $G'$ is the conductance of the system without the
third tip c) is the LDOS. d) (blue line) Conductance $G'$ of system
as a function of $d_{\mathrm{tip}}$ without third tip. (red line)
Absolute value of Pearson correlation $r$ between (a) and (c) images
calculated from Eq. \ref{eq:r}.}
\end{figure*}

\begin{figure}[h]
\begin{centering}
\includegraphics[width=0.4\paperwidth]{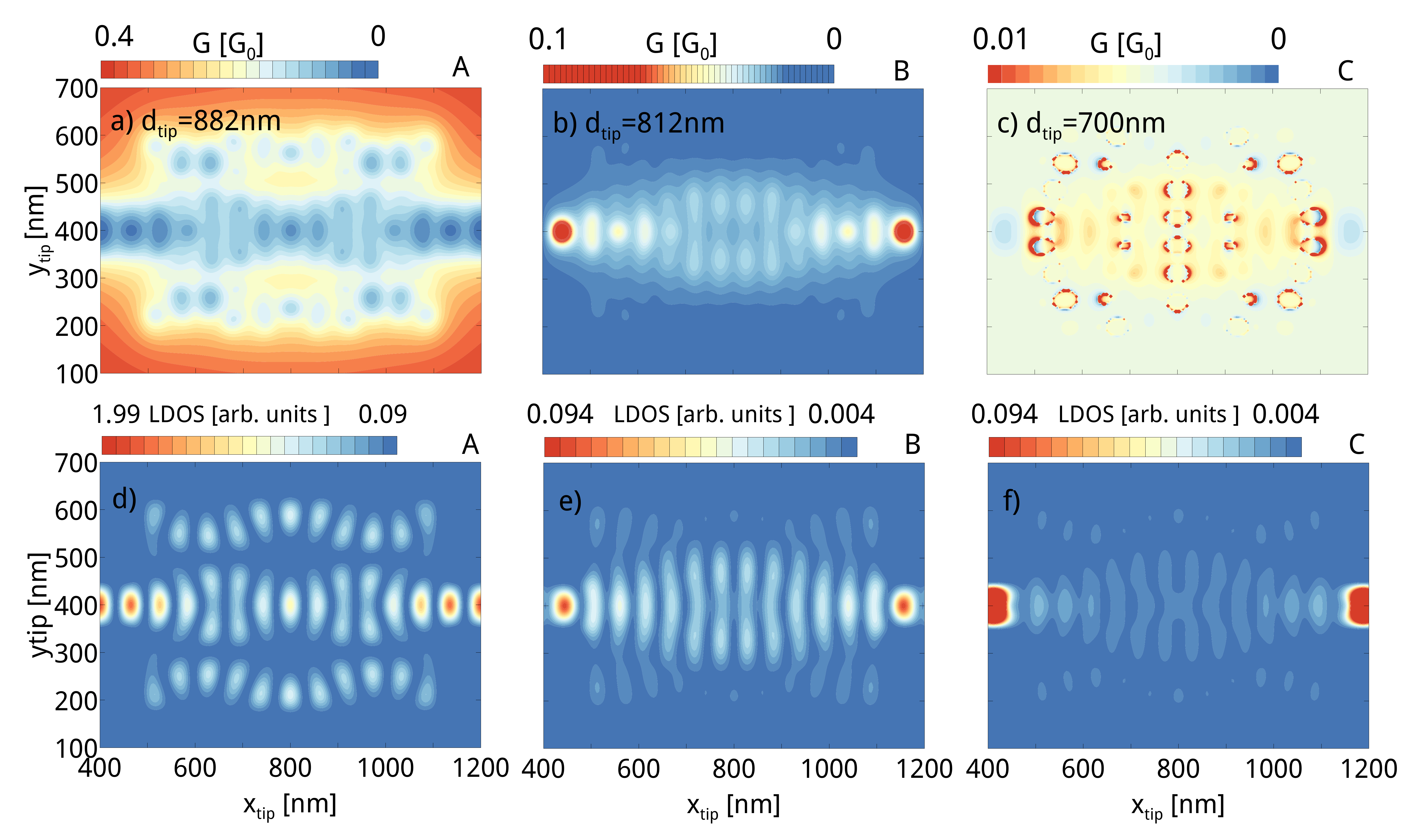}
\par\end{centering}

\caption{\label{fig:korelacjeSGM} a-c) The SGM image d-f) and LDOS obtained
for point \textbf{A,B} and \textbf{C} in Fig. \ref{fig:korelacje}(d).
Point \textbf{A} corresponds to the symmetric resonance induced by
the double tip system in work point where the correlation $r$ is
negative. Point \textbf{B} corresponds to the antisymmetric resonance
with positive correlation $r$. Point \textbf{c} corresponds to a
zero of $T$ in the absence of the tip.}
\end{figure}

In Fig. \ref{fig:korelacje}(a) we show the conductance of the system
depicted of Fig. \ref{fig:wnekaSchemat} as a function of distance
between the two tips inside the channels $d_{tip}$ (see Fig. \ref{fig:rezonanseWtip})
and position of the third tip $x_{tip}$ along the red line in Fig.
\ref{fig:wnekaSchemat}. We observe a number of resonances for a sequence
of the tip positions. The resonances appear in pairs, which have LDOS
with a minimum or a maximum at the symmetry axis of the device {[}Fig.
\ref{fig:korelacje}(c){]} with either a maximum or a zero of LDOS
at the symmetry axis $x_{tip}=800$ nm. The $d_{tip}$ distance which
corresponds to resonances can be deduced from the conductance dependence
of the cavity -- double tip system (without the third tip) which is
plotted in Fig. \ref{fig:korelacje}(d). In Fig. \ref{fig:korelacje}(b)
we present the deviation $\Delta G$ from the conductance $G'$ of
the system without the third tip. Note that $\Delta G$ changes sign
in the vicinity of the resonance induced by double tip (see the blue
curve in Fig. \ref{fig:korelacje}(d)), which will lead to a negative correlation of between
the LDOS and G map\cite{KolasinskiNJP}. In order to compare the conductance
obtained from scan with third tip $G_{\mathrm{scan}}\equiv G(x_{\mathrm{tip}},d_{\mathrm{tip}})$
with the LDOS on that line $\mathrm{L_{scan}}=\mathrm{LDOS}(x_{\mathrm{tip}},d_{\mathrm{tip}})$
(for given value of $d_{\mathrm{tip}}$) we calculate the Pearson
correlation $r\equiv r(d_{\mathrm{tip}})$
\begin{equation}
r=\frac{\int(G_{\mathrm{scan}}-\langle G_{\mathrm{scan}}\rangle)(\mathrm{L_{scan}}-\langle\mathrm{L_{scan}}\rangle)dx_{\mathrm{tip}}}{\Delta_{\mathrm{L}}\sigma(G_{\mathrm{scan}})\sigma(L_{scan})},\label{eq:r}
\end{equation}
where $\ensuremath{\langle a\rangle=\frac{1}{\Delta_{\mathrm{L}}}\int a(x_{\mathrm{tip}})dx_{\mathrm{tip}}}$
is the average value of $a$, $\ensuremath{\sigma^{2}(a)=\frac{1}{\Delta_{\mathrm{L}}}\int(a(x_{\mathrm{tip}})-\langle a\rangle)^{2}dx_{\mathrm{tip}}}$
is standard deviation of $a$ and $\Delta_{\mathrm{L}}=800$nm is
the length of the scan along red line in Fig. \ref{fig:wnekaSchemat}.
Before the calculation of $r$ both functions $G_{\mathrm{scan}}$
and $L_{scan}$ were normalized to range from 0 to 1. The absolute
value of the correlation $r$ is plotted in Fig. \ref{fig:korelacje}(d).
Note that $r$ increases to about 0.9 near the induced resonances
which means that obtained G map well resolve the LDOS inside the cavity.
The dips in $r(d_{tip})$ dependence result from the fact that the
correlation coefficient changes sign at the resonances \cite{KolasinskiNJP}.
In Fig. \ref{fig:korelacjeSGM} we present the SGM images and corresponding
LDOS for points A, B and C of Fig. \ref{fig:korelacje}(d). Summarizing,
the usage of multiple tips allows for 1) tuning the system to a resonance
and then 2) an accurate read-out of the local density of states.

For evaluation of the LDOS we normalized the scattering densities
to the unity within the computational box. For off-resonant conditions,
where the scattering density is large only inside the input and output
channels, we thus obtain a strong reduction of LDOS within the cavity.
This normalization approach gives a close correspondence of LDOS with
SGM maps for varied $d_{tip}$ and $x_{tip}$ {[}cf. Fig. \ref{fig:korelacje}{]}.
For fixed values of these parameters, the Pearson correlation coefficients
between the maps is not affected by the LDOS normalization.

\section{Tip as a part of the scattering system}

The scanning gate microscope tip can be used
as a part of the scattering system, with the other tip gathering the conductance maps.
In particular, the tip potential placed near the exit of the quantum point contact
has been used to form a quantum Hall interferometer \cite{enslin}.
A similar quantum Hall system including a potential defect spontaneously formed at the exit from the QPC constriction
was studied with the scanning gate microscopy in Ref. \cite{martins}

\begin{figure}[h]
\begin{centering}
\includegraphics[width=0.4\paperwidth]{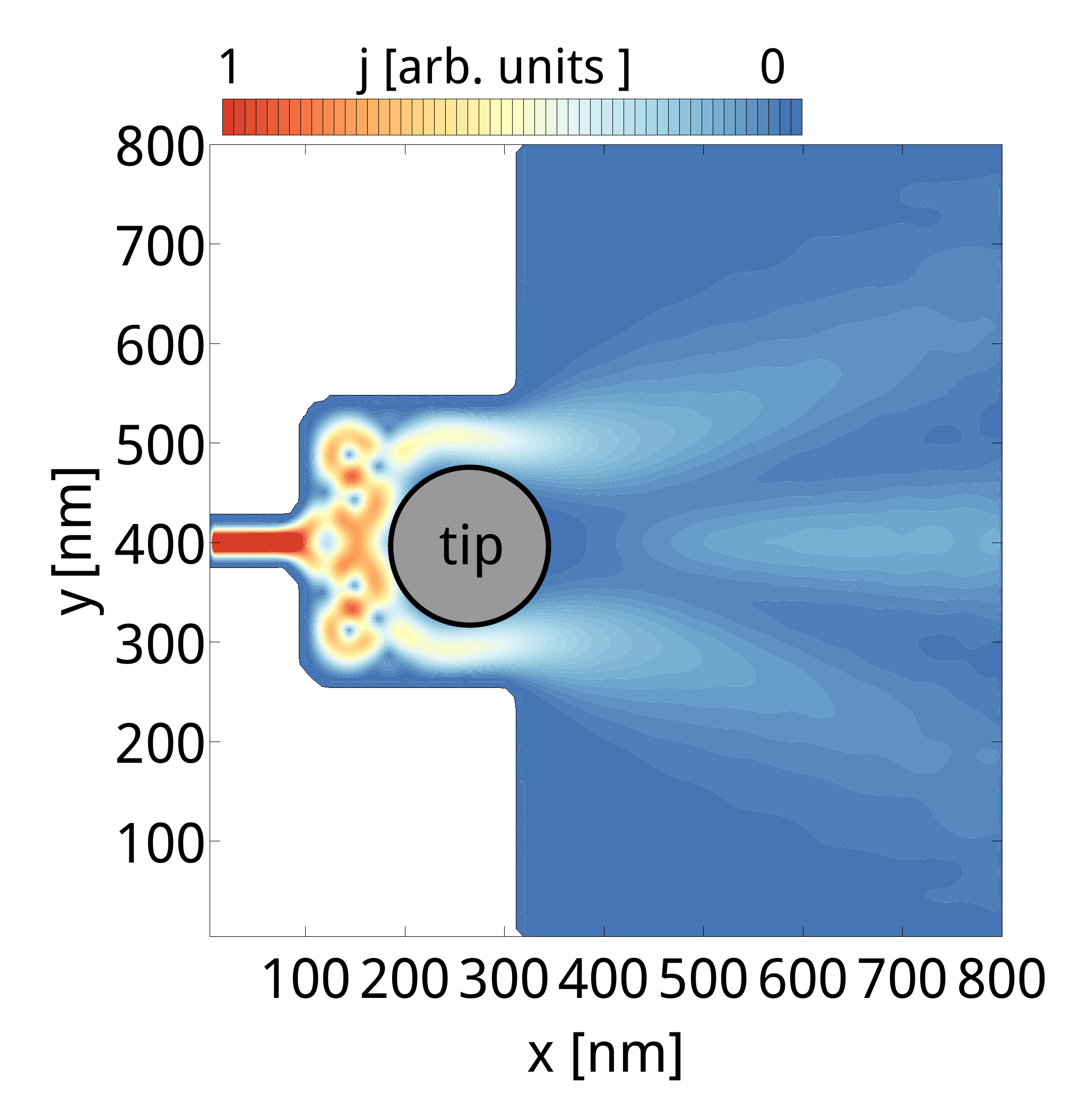}
\par\end{centering}
\caption{\label{qqhi}  Schematics of system with a quantum point contact and the tip potential for observation
of the Young interference and the calculated scattering density current $|j(x,y)|$ including the double-slit interference pattern.
The applied tip potential  parameters: $w_{\mathrm{tip}}=50$nm and $U_{\mathrm{tip}}=10$meV}
\end{figure}

The SGM with a double tip system can be used for observation of the Young interference \cite{qhi} for a set-up depicted in Fig. \ref{qqhi}.
The SGM signal was predicted \cite{qhi} to contain signatures of the double-slit interference
for low-energy transport. The proposed \cite{qhi} set-up requires a beam splitter -- a central obstacle for the electron flow,
which may be difficult to introduce in form of a fixed gate. A floating one -- introduced by the tip -- is
a possible option. In Fig. \ref{qqhi} we plotted a system
with the QPC slit entering a small open cavity. The potential of the tip (schematically shown with the gray circle) was introduced near the exit
of the cavity to an unconfined half-plane. The scattering density current $|j(x,y)|$ plot in Fig. \ref{qqhi} shows
formation of the interference pattern characteristic for the Young interference \cite{qhi}.
The other tip used as an electron flow detector could then be used for readout
of the double-slit interference in the SGM conductance images.

\section{Conclusions}

We have studied numerically the possible applications of the multitip
scanning gate microscopy on the systems with 2DEG. We found that the
measurements using a pair of tips allows for 1) readout of the Fermi
wavelength 2) detection of the potential defects with indication of
their position and profile, 3) filtering of specific transverse modes
4) tuning the system into a resonance that allows for a reliable detection
of the local density of states with a third tip.

\textbf{Acknowledgments} This work was supported by National Science
Centre according to decision DEC-2012/05/B/ST3/03290, and by PL-Grid
Infrastructure. The first author is supported by the scholarship of
Krakow Smoluchowski Scientific Consortium from the funding for National
Leading Reserch Centre by Ministry of Science and Higher Education
(Poland).

\end{document}